\documentclass[11pt]{article}

\usepackage[letterpaper,margin=0.82in]{geometry}
\usepackage[T1]{fontenc}
\usepackage{newtxtext,newtxmath}
\usepackage{microtype}
\usepackage{parskip}
\usepackage{enumitem}
\usepackage{xcolor}
\usepackage{lineno}
\usepackage{titlesec}
\usepackage{comment}
\usepackage{fancyhdr}
\usepackage[hidelinks]{hyperref}
\usepackage[
  backend=biber,
  style=numeric,
  sorting=none
]{biblatex}

\DeclareFieldFormat[
  article,
  book,
  inproceedings,
  misc,
  report,
  techreport
]{title}{\mkbibemph{#1}}

\definecolor{sectionblue}{RGB}{35,72,112}

\titleformat{\section}
  {\large\bfseries\color{sectionblue}}
  {}{0pt}{}
\titlespacing*{\section}{0pt}{1.0em}{0.35em}

\setlist{nosep,leftmargin=1.4em}

\fancypagestyle{firstpage}{
  \fancyhf{}
  \lhead{\small U.S. Muon Collider Collaboration}
  \rhead{\small Input to the SCAC Facilities Subcommittee}
  \cfoot{\parbox{0.88\textwidth}{\footnotesize * The U.S. Muon Collider Collaboration (USMCC) includes nearly 300 scientists and engineers from over 80 institutions, bringing together accelerator scientists, experimentalists and theorists from U.S. universities and national laboratories, including a substantial and growing community of students, postdoctoral researchers and early career scientists.}\\[-0.15em]\small\thepage}

}

\begin{document}
%\linenumbers
\begin{center}
{\Large\bfseries Input to the SCAC Facilities Subcommittee from the U.S. Muon Collider Collaboration}

\vspace{0.35em}

{\normalsize
S.~Jindariani$^{1}$, T.~Holmes$^{2}$, P.~Meade$^{3}$ K.~DiPetrillo$^{4}$,
K.~Kennedy$^{5}$, L.~Lee$^{2}$, S.~Pagan Griso$^{6}$, D.~Stratakis$^{1}$
\\
\bf On behalf of the U.S. Muon Collider Collaboration$^{*}$}

{\small

$^{1}$Fermi National Accelerator Laboratory, Batavia, Illinois, USA\\

$^{2}$University of Tennessee, Knoxville, Tennessee, USA\\

$^{3}$Stony Brook University, Stony Brook, New York, USA\\

$^{4}$University of Chicago, Chicago, Illinois, USA\\

$^{5}$Tufts University, Medford, Massachusetts, USA\\

$^{6}$Lawrence Berkeley National Laboratory, Berkeley, California, USA

}
\end{center}

\vspace{0.5em}
\thispagestyle{firstpage}
The U.S. Muon Collider Collaboration welcomes the opportunity to contribute to SCAC’s facilities planning. A Muon Collider offers a distinctive path to the 10 TeV parton-energy scale, a central long-term goal of U.S. particle physics. P5 recommended targeted R\&D to establish its feasibility, with a demonstrator as an important milestone~\cite{P5}. The National Academies subsequently called for an immediate national Muon Collider R\&D program, in partnership with international efforts, to demonstrate the key technologies and their integration~\cite{NAS2025}.

SCAC’s 2026–2036 planning period is critical to this longer-term vision~\cite{SCACCharge,DOEFacilities,SCACDrell}. The major facilities that the United States may need in the 2040s cannot enter the facilities pipeline in the 2040s: their scientific concepts, technologies, workforce and predecessor facilities must be developed during the preceding decade. The near-term priority is therefore not collider construction, but the focused R\&D and smaller-scale demonstrators needed to establish the technical basis for the next stage.

We propose a coordinated, staged program beginning with micro-demonstrators: tests of components and subsystems distributed across U.S. laboratories and universities, matched to existing capabilities and complementary to other accelerator and magnet R\&D. These tests would reduce technical risk while advancing technologies with broader scientific and industrial applications. In parallel, the program would develop the design of a macro-demonstrator: an integrated facility for producing and cooling muon beams. Its configuration and potential physics program, including neutrino, dark matter, flavor and nuclear physics, would be assessed as part of this work. Results from the micro-demonstrators would establish the basis for an informed decision about this next-stage facility.

We encourage SCAC to include this R\&D and demonstrator pathway within its facilities planning framework, alongside major facilities approaching construction. Building on existing U.S. infrastructure, such a program would strengthen the accelerator workforce, advance capabilities across the Office of Science, and preserve the option of U.S. leadership and future hosting at the energy frontier. The following sections address the key considerations guiding SCAC’s facilities planning.

\section{Physics opportunities}

A muon collider offers a distinctive route to the 10 TeV parton-energy scale. A 100 TeV proton collider such as FCC-hh and a 10 TeV muon collider may have similar reach for selected benchmarks, but their collisions, and therefore the physics they can uncover, are substantially different.

Proton colliders are robust discovery machines for particles carrying color, the charge of the strong force, and can reach important electroweak targets. A muon collider is inherently a high-energy electroweak collider, particularly suited to central post-LHC questions: What determines our universe’s vacuum? Is the Higgs elementary or composite, and is it the only scalar? What stabilizes the electroweak scale? Does the Higgs connect ordinary matter to dark matter or hidden sectors?

%Understanding our vacuum requires more than measuring how the Higgs couples to itself, based on interactions between three or more Higgs bosons. Measurements of the Higgs boson constrain only part of the scalar potential. Additional scalar fields can change the structure of our vacuum with only small effects on the observed Higgs boson itself, making it essential to search directly for these new particles.

Understanding our vacuum requires more than Higgs self-coupling measurements. Measurements of the known Higgs constrain only part of the scalar potential. Additional scalar fields can change our vacuum  with only small effects on the observed Higgs boson itself, making direct searches essential.

A muon collider combines direct production of new electroweak states, including additional Higgs-like and Higgs-portal particles difficult to isolate at proton colliders, with precision studies of abundant Higgs, W, Z, and top particles.  
This gives a robust tool both to search for new physics below the 10 TeV scale, motivated by our post-LHC questions, and to reach well beyond 10 TeV through the quantum imprints of heavier states.
In particular, the contributions of higher-dimensional quantum operators grow with collision energy and can be probed in an unprecedented fashion with a muon collider.  The combination of many muon collisions near the machine’s full energy with the large center of mass energy can probe new interactions associated with scales approaching 100 TeV.

A composite Higgs illustrates this power. Internal structure could produce correlated changes in high-energy production of Higgs pairs, top-quark pairs, and other quark and lepton pairs. In benchmark scenarios, these measurements can test compositeness scales well beyond FCC-hh’s projected reach. This illustrates how the muon collider is a two-for-one machine, providing both precision and discovery: observing new physics indirectly and, when the responsible particles are accessible, producing them directly to determine their properties.

Beyond the Higgs, the collider can explore electroweak dark matter, including candidates that underground experiments may hint at but cannot fully identify. It also brings a complementary perspective on flavor, the unexplained pattern of particle families, masses, and interactions. Low-energy experiments and high-energy muon collisions can probe the same quantum operators. Higher energy offers a different lever arm from larger low-energy datasets, potentially connecting low-energy effects or flavor-dependent Higgs couplings to direct discoveries.

The collider would anchor a broader research complex. Its powerful proton source could support rare-kaon, precision-muon, fixed-target, and hidden-sector experiments. Muon production and acceleration offer a path to a neutrino factory building on DUNE’s far-detector infrastructure, while higher-energy beams could enable precision tau-neutrino studies. Muon–nucleus scattering could probe quark and gluon behavior, including gluon saturation; neutrino deep-inelastic scattering could map nuclear structure, test electroweak interactions, and measure quark-flavor mixing.

The result is a robust, multipurpose program centered on electroweak symmetry breaking, the structure of our vacuum, and connections to dark matter, flavor, neutrinos, and nuclear physics.

\section{From R\&D to demonstration}

Substantial progress has been made on Muon Collider designs and technologies over the past several years~\cite{ESPPU,Princeton2024,InternationalMuonCollider:2025sys}. The International Muon Collider Collaboration (IMCC), which includes much of the U.S. community, has developed increasingly detailed accelerator and detector concepts. Muon cooling remains the biggest risk in the collider's luminosity performance. The principle of ionization cooling of muons has been experimentally demonstrated by MICE~\cite{MICE} and g-2 experiments~\cite{Bradley:IPAC2018-TUPMK015}, but the feasibility of a practical implementation of a Muon Collider-like cooling channel still needs to be established. Such a channel will require many cooling stages incorporating high-field solenoids, RF cavities and absorbers, operating together under realistic engineering conditions, achieving the required cooling while maintaining sufficient muon transmission~\cite{P5Submission,Princeton2024}.

There has also been important progress in the overall accelerator design, detector concepts, high-field magnets, RF, targetry and simulation tools~\cite{ESPPU,Princeton2024,AI4MuC}. No fundamental showstoppers have been identified, but major engineering and integration challenges remain. Muons must be produced using a high-power proton source, captured, cooled by approximately six orders of magnitude in six dimensional phase space, rapidly accelerated and collided before they decay. This requires high-power target and capture systems, high-field superconducting solenoids, high-gradient RF operating in strong magnetic fields, fast ramping magnets and associated power systems. These must be designed with consideration of beam dynamics, radiation mitigation and the machine detector interface. On the detector side, beam induced backgrounds require precision timing, high granularity, radiation tolerant electronics and significant advances in simulation and reconstruction~\cite{ESPPU,Princeton2024}.

University resources, private foundations (including the Simons Foundation, Research Corporation for Science Advancement, and Sloan Foundation), and limited targeted investments at national laboratories have sustained U.S. Muon Collider studies, but cannot support the coordinated accelerator and technology R\&D needed this decade. The absence of dedicated DOE support also limits participation by national-laboratory scientists and engineers, despite their essential expertise and infrastructure for addressing the technical challenges.

A valuable next step would be a focused, staged R\&D and demonstrator effort aimed at refining the overall accelerator design while developing and testing the most critical technologies. We propose a distributed and staged approach, starting with small-scale demonstrators and progressing toward increasingly integrated systems. These micro-demonstrators can provide rapid feedback, benchmark simulations, support choices among competing technologies and reduce technical risk before larger commitments are made. They should address magnets, RF cavities and power sources, target and capture systems, fast ramping magnets and their power systems, detector technologies and integrated cooling cell prototypes~\cite{P5Submission,Princeton2024}.

\subsection*{Developing the concept for an integrated demonstrator facility}
Design of the macro-demonstrator concept should proceed in parallel with component R\&D. Earlier planning focused primarily on an ionization cooling demonstrator consisting of a sequence of cooling cells, together with a muon source and diagnostics~\cite{P5Submission,Princeton2024}. Such a facility was intended to demonstrate significant cooling and integration of the technologies required for a realistic cooling channel; it was not designed as a high intensity facility with a significant particle physics program of its own.

A higher intensity demonstrator could address a broader set of Muon Collider challenges. Such a facility could potentially demonstrate important elements of the front end, including proton beam delivery, target and capture, muon production and cooling, while providing scientifically interesting capabilities in its own right. The intense muon and neutrino beams from this facility have HEP and NP applications, as well as many synergies with muon-catalyzed fusion~\cite{Kalow:2026MuFusE}.
%One possibility being investigated is whether such a demonstrator could provide capabilities of a neutrino factory.
Its exact configuration, intensity and scientific scope are not yet established. Developing these concepts and determining whether a compelling physics program can be incorporated should be part of the R\&D effort during the coming years.

During 2026--2036, micro-demonstrator results should inform parallel design of the macro-demonstrator, establishing its scope, technical feasibility, scientific opportunities, cost and schedule for an informed U.S. decision about the next stage. The macro-demonstrator would, in turn, inform collider design.

\subsection*{Potential for maintaining U.S. leadership and accelerating scientific discovery}
\begin{comment}
A Muon Collider offers the possibility of a future U.S. hosted facility at the energy frontier, while providing scientific capabilities complementary to those of a future hadron collider. Both P5 and the National Academies have recognized this potential~\cite{P5,NAS2025}. Maintaining that option requires significant U.S. participation well before a construction decision. The design and technology development phase is where much of the expertise and leadership needed for a future demonstrator, and ultimately a collider, will be established.

Muon Collider R\&D also advances capabilities before construction of a collider. The program requires advances in superconducting magnets, high power targets, RF, beam dynamics, detector instrumentation, radiation tolerant electronics and computing. Artificial intelligence could accelerate this work by allowing experts to explore much larger design spaces and better optimize strongly coupled systems. The AI4MuC workshop identified cooling channel optimization, target and capture design, multiphysics optimization of magnets and RF, detector optimization and beam induced background simulation as areas where AI and machine learning can have significant impact~\cite{AI4MuC}. This provides a natural connection to broader DOE efforts in AI enabled scientific discovery. Longer term, Muon Collider problems may also provide useful test cases for emerging computational approaches, including quantum computing, although no specific quantum advantage has yet been demonstrated.
\end{comment}

A Muon Collider offers the possibility of a future U.S. hosted facility at the energy frontier, while providing scientific capabilities complementary to those of a future hadron collider. The HEPAP International Benchmarking Subpanel emphasized that U.S. leadership in particle physics depends on maintaining strong domestic capabilities, experience and infrastructure, and that hosting major facilities is an important foundation for international leadership and partnership~\cite{Benchmarking}. Maintaining that option requires significant U.S. participation well before a construction decision. The design and technology development phase is where much of the expertise and leadership needed for a future demonstrator, and ultimately a collider, will be established.

Artificial intelligence could accelerate Muon Collider R\&D by allowing experts to explore much larger design spaces and better optimize strongly coupled systems. The AI4MuC workshop identified cooling channel optimization, target and capture design, multiphysics optimization of magnets and RF, detector optimization and beam induced background simulation as areas where AI and machine learning can have significant impact~\cite{AI4MuC}. This provides a natural connection to broader DOE efforts in AI enabled scientific discovery. Longer term, Muon Collider problems may also provide useful test cases for emerging computational approaches, including quantum computing, although specific quantum advantage has not yet been demonstrated.

\subsection*{Role in critical workforce development}
\begin{comment}
The shortage of accelerator scientists has been identified as a critical threat to the future of U.S. particle physics, resulting in a task force specifically aimed at this problem~\cite{NAS2025, Clarke:2026jri}. Maintaining this expertise requires more than training programs: there must be challenging projects that attract students and young scientists and allow them to develop technologies that have not been built before. A Muon Collider provides exactly this environment. The work spans accelerator physics, superconducting magnets, RF, high power targets, detector instrumentation, AI and computing, and brings together expertise from national laboratories and universities.

The 2024 Princeton workshop found substantial interest and expertise in the university community, but also identified the need for mechanisms that allow university researchers and students to work closely with accelerator experts at the national laboratories~\cite{Princeton2024}. A focused national program would provide such a mechanism, allow experienced accelerator scientists to train the next generation while this expertise is still available, and develop a workforce needed not only for a future Muon Collider but for the broader DOE accelerator program.
A sustained and coherent R\&D program is important for preserving and building upon the specialized knowledge and expertise already developed, ensuring that each stage of the program benefits from the experience and results of the previous one.
\end{comment}
The accelerator workforce has been identified as a critical area for sustaining the future of U.S. particle physics~\cite{NAS2025, Clarke:2026jri}. Training programs alone are insufficient: challenging projects must attract students and young scientists and enable them to develop new technologies. Muon Collider R\&D provides this environment by bringing together national-laboratory and university expertise.

The 2024 Princeton workshop found substantial interest and expertise in the university community, but also identified the need for mechanisms that allow university researchers and students to work closely with accelerator experts at the national laboratories~\cite{Princeton2024}. A coordinated national R\&D effort could provide such a mechanism, enabling experienced accelerator scientists to train the next generation while this expertise remains available not only for a future Muon Collider but for the broader DOE accelerator program.

\subsection*{Ability to leverage existing capabilities}

A U.S. Muon Collider R\&D program can make extensive use of existing DOE capabilities rather than requiring construction of a major new facility during the coming decade. The National Lab Accelerator Study Group (NLASG)~\cite{NationalLabMuonColliderStudy} identified substantial existing capabilities and important synergies between Muon Collider R\&D and other Office of Science programs. Existing facilities can support component tests, beam studies and many of the proposed small scale demonstrators. Important test capabilities that should be considered include a combined RF and magnetic field test stand, an integrated cooling cell test facility, and a high power target test station. Maintaining test-beam capabilities in the U.S. is important not only for the Muon Collider R\&D program, but also for supporting a broad portfolio of accelerator and detector R\&D across HEP and Nuclear Physics. The program should make use of existing facilities wherever possible and develop dedicated test stands and test facilities only where they are needed. The integrated demonstrator facility can build on existing proton driver facilities within the U.S. National Laboratories.

\subsection*{Synergies and partnerships}

Many of the required technologies have strong connections to other DOE programs. High-field HTS magnets required for muon cooling have synergies with fusion and high-field science. High power target R\&D overlaps with challenges faced by LBNF/DUNE, SNS and other intense beam facilities. RF, fast ramping magnets and efficient power systems have broader accelerator applications, affecting not only scientific accelerator facilities but also medical applications like proton therapy. Precision timing, radiation hard electronics and intelligent detectors are important across HEP, but also in medical imaging, fusion technologies, and nuclear security~\cite{Princeton2024,AI4MuC}. These connections also create opportunities for partnerships among DOE laboratories, universities and industry as technologies progress from concepts toward engineered systems.

International partnership is also essential. IMCC has established the framework for the international design effort and U.S. scientists are already significant contributors~\cite{ESPPU,Princeton2024}. The 2026 European Strategy for Particle Physics recommends continued development of technologies underpinning high brightness muon beams and explicitly calls for cooperation with international partners, including the U.S. Muon Collider R\&D initiative~\cite{ESPPU2026}. 

%Activities during the coming decade could establish technical readiness, define the demonstrator's scope and scientific opportunities, develop credible cost and schedule estimates, identify opportunities to leverage existing U.S. infrastructure, and clarify appropriate U.S. and international roles.

\subsection*{National and economic security}

The primary motivation for a Muon Collider R\&D program is scientific, but several required technologies, including high-field superconducting magnets and advanced superconductors, high-power RF and targets, radiation tolerant microelectronics, advanced power systems and AI enabled scientific computing, are strategically important U.S. capabilities. Maintaining domestic expertise and industrial capability in these areas contributes to U.S. technological competitiveness. More broadly, preserving the scientific and engineering workforce capable of designing and constructing large accelerator systems is itself a national capability that is difficult and slow to rebuild once lost.

\subsection*{Risks to cost and schedule}

\begin{comment}
A future Muon Collider demonstrator is not yet sufficiently mature for a credible construction cost and schedule. The appropriate way to reduce these uncertainties is a staged R\&D program: design, prototype and test critical components; progress toward subsystem and integration tests; and mature the overall accelerator and detector designs sufficiently to establish realistic technical requirements, infrastructure needs, siting options and cost ranges. This is consistent with SCAC's own emphasis on additional R\&D before construction when needed to establish scope, reduce technical risk and limit future cost and schedule growth~\cite{SCACDrell}. The R\&D program therefore directly reduces the risk of a future facility.
\end{comment}
A future Muon Collider demonstrator is not yet sufficiently mature for a credible construction cost and schedule. The appropriate way to reduce these uncertainties is a staged R\&D program: design, prototype and test critical components; progress toward subsystem and integration tests; and mature the overall accelerator and detector designs sufficiently to establish realistic technical requirements, infrastructure needs, siting options and cost ranges. This is consistent with SCAC's own emphasis on additional R\&D before construction when needed to establish scope, reduce technical risk and limit future cost and schedule growth~\cite{SCACDrell}.

\section{The ask to SCAC}

%P5 and the National Academies have identified the Muon Collider as one of the most promising options for the long term future of U.S. particle physics~\cite{P5,NAS2025}. The International Benchmarking Subpanel emphasized the importance of retaining the domestic capabilities required for the United States to lead in international particle physics~\cite{Benchmarking}, while the 2026 European Strategy explicitly anticipates cooperation with the U.S. initiative on muon beam R\&D~\cite{ESPPU2026}.

%We encourage SCAC to consider the current ten year planning horizon in the context of the longer timescale required to develop major new scientific facilities. The portfolio created during 2026 to 2036 should not only advance facilities approaching construction, but also leave the United States with compelling and technically credible options for the decade that follows. A healthy facilities ecosystem requires both major construction projects and a pipeline of R\&D facilities, test stands and demonstrators that enable the next generation.

For 2026–2036, we ask SCAC to recognize the missing Muon Collider R\&D and demonstrator pathway as a gap in the facilities portfolio. The priority is a coordinated program of accelerator and detector design, component demonstrators and essential test facilities, with parallel design of an integrated muon-production and cooling demonstrator. This work should establish its technical feasibility, scientific opportunities, scope, cost, schedule, infrastructure needs and domestic and international roles before a construction decision. Building on existing U.S. capabilities, this program would advance technologies across the Office of Science, develop the accelerator workforce and preserve the option of U.S. leadership and future hosting at the energy frontier.

\printbibliography

\end{document}